\documentclass[pra, twocolumn, superscriptaddress, amsmath, amssymb, aps,longbibliography]{revtex4-2}

\usepackage[pdftex]{graphicx}
\usepackage{svg}
\usepackage{dcolumn}
\usepackage{bm}
\usepackage{dsfont}
\usepackage{physics}
\usepackage{mathrsfs}
\usepackage[colorlinks, breaklinks]{hyperref}
\usepackage{amsmath}
\usepackage[english]{babel}
\usepackage{booktabs}
\usepackage{amssymb}
\usepackage{type1cm}
\usepackage[caption=false]{subfig}
\usepackage{braket}
\usepackage{xcolor}
\usepackage{comment}
\usepackage{algorithm}
\usepackage{algpseudocode}
\usepackage{tikz}
\usetikzlibrary{quantikz2}

\begin{document}

\title{Resource Estimation for Delayed Choice Quantum Entanglement Based Sneakernet Networks Using Neutral Atom qLDPC Memories}

\author{S. Srikara}
\affiliation{Center for Quantum Software and Information, University of Technology Sydney, NSW 2007, Australia.}

\author{Andrew D. Greentree}
\affiliation{Australian Research Council Centre of Excellence for Nanoscale BioPhotonics, School of Science, RMIT
University, Melbourne, VIC 3001, Australia.}

\author{Simon J. Devitt}
\affiliation{Center for Quantum Software and Information, University of Technology Sydney, NSW 2007, Australia.}
\affiliation{InstituteQ, Aalto University, 02150 Espoo, Finland.}

\begin{abstract}
Quantum Entanglement is a vital phenomenon required for realizing secure quantum networks, so much that distributed entanglement can be re-imagined as a commodity which can be traded to enable and maintain these networks. We explore the idea of commercializing entanglement-based cryptography and future applications where advanced quantum memory systems support less advanced users. We design a sneakernet-based quantum communication network with a central party connecting the users through delayed-choice quantum entanglement swapping, using quantum Low-Density-Parity-Check (qLDPC) encoded qubits on neutral atoms. Our analysis compares this approach with traditional surface codes, demonstrating that qLDPC codes offer superior scaling in terms of resource efficiency and logical qubit count. We show that with near-term attainable patch sizes, one can attain medium-to-high fidelity correlations, motivating further research towards the long-term realization of large-scale commercial quantum networks.
\end{abstract}

\maketitle

\section{Introduction}

The second Quantum Revolution refers to a future technological period when all the features of quantum mechanics are harnessed for information processing devices, expanding beyond the limited application of quantum mechanics in current computers, communication systems, and sensors \cite{dowling2003quantum,deutsch2020harnessing,jaeger2018second,atzori2019second,rohde2021quantum}. The ultimate achievement of this revolution is embodied by quantum computers that are capable of operating on a vast scale, with the ability to tolerate faults and fix errors. These computers would need millions or more logical qubits \cite{riel2021quantum} to tackle complex cryptographic tasks. While the creation of such devices remains in its early stages, there are compelling grounds to anticipate that scalable quantum computers will become tangible at some point within this century \cite{gambetta2020ibm}. 

While there is reason to be hopeful about the progress of large-scale quantum computers, it remains uncertain if quantum computers for everyday consumers, such as a quantum smartphone, will ever become feasible. This refers to a compact and powerful computational device that is connected to a quantum network, possibly on a global scale. Possible constraints that could prevent the development of quantum smartphones include the requirement for extremely low temperatures to cool down qubits \cite{gardiner2015quantum}, or the necessity for optical communication channels \cite{o2007optical} instead of microwave ones to connect to the quantum internet, the vulnerability of qubits to mechanical vibrations and the impact of the Earth's magnetic field \cite{korber2018decoherence}. Suppose we accept the possibility of large-scale quantum computers and a quantum internet. In that case, it is logical to enquire how a customer, who has no access to a quantum memory, but only has access to conventional gear such as classical detectors, classical memory, and classical computing, can participate in and perform complex quantum protocols.

One way is to do this by the use of distributed quantum entanglement \cite{perseguers2013distribution}. Quantum entanglement is a pivotal phenomenon that forms the basis of various protocols in quantum communications and quantum computing \cite{quantuminternet, nielsen2001quantum, entanglement1,wehner2018quantum,kimble2008quantum,cacciapuoti2019quantum}. Due to it being a major primitive for most quantum tasks, it would be natural to consider quantum entanglement as a tradeable commodity, such that users can purchase it to establish connections with other users in the network to perform quantum tasks such as quantum key distribution, distributed quantum computing, quantum authentication, etc. Commoditizing entanglement is an essential step toward large-scale commercial quantum networks. 

To better demonstrate the above-mentioned ideas of entanglement commercialization and quantum access to common users with classical hardware, we design and analyse an entanglement one-time-pad QKD \cite{sharma2019survey} based quantum network model networked using delayed-choice entanglement swapping \cite{peres2000delayed}. In this study, we demonstrate that delayed-choice entanglement swapping enables customers, who only own classical hardware, to use the offerings of a hypothetical future quantum memory company for the purpose of generating secure quantum keys. These keys are based on the E91 protocol \cite{ekert1991quantum}. Our design employs a central party Charlie, who generates Bell-pairs, loads the halves separately onto different qLDPC codes, and distributes one set of the halves via sneakernet \cite{devitt2016high,zhong2015optically} to multiple endpoints where it is immediately measured upon arrival while keeping the other set of halves to himself on which he later performs Bell measurements to generate correlations between the endpoint measurements of any pair of users.  This enables the users to execute this task in a versatile manner by pre-purchasing their key bits prior to selecting their communication partners. This protocol is beneficial for creating secure connections between networks of devices that have minimal hardware needs but can interact with a single, advanced quantum node or several nodes. 

We design our protocol on quantum Hyper Graph Product (HGP) codes \cite{tillich2013quantum}, which are a class of Low-Density Parity-Check (qLDPC) codes \cite{breuckmann2021quantum} for quantum error correction \cite{lidar2013quantum}, implemented on a neutral atom architecture \cite{henriet2020quantum,xu2024constant}. These codes offer significant advantages over traditional surface codes \cite{fowler2012surface} in terms of encoding rate and resource efficiency \cite{bravyi2024high,xu2024constant}. Our analysis provides a detailed comparison between qLDPC and surface codes, demonstrating that qLDPC codes allow us to achieve higher logical qubit counts with fewer physical qubits as they have constant encoding rates regardless of patch size. This property potentially reduces the hardware requirements and improves the scalability of our proposed system.
We have selected neutral atoms as our physical platform, as their ability to shuttle and rearrange within the lattice enables selective qubit unloading and efficient syndrome extraction \cite{xu2024constant}. This capability also allows us to use only one set of ancilla qubits and a single surface code patch, positioned adjacent to the qLDPC code block, for both loading and unloading operations via teleportation \cite{xu2024constant}.

The foundation of our study is based on the research conducted on consumer-level quantum key distribution \cite{zhang2018large, lowndes2021low, duligall2006low}, as well as the research on quantum sneakernet \cite{devitt2016high}, which explores the spread of entanglement across long distances with high bandwidth and low latency. While our primary focus is on secure communication protocols, it is feasible to adapt our methodology for many applications such as quantum voting \cite{vaccaro2007quantum}, clock synchronization \cite{giovannetti2001quantum}, and reference frame alignment \cite{bagan2004quantum}.

This study is organized as follows: In section \ref{sec:dces}, we briefly describe the phenomenon of delayed choice entanglement swapping. In section \ref{sec:dcsetup}, we describe our design. Section \ref{sec:dcresourceestimation} derives the closed-form mathematical expressions necessary for our resource estimations. In section \ref{sec:results} we discuss our findings, and finally conclude our study in section \ref{sec:conclusion}. Link to the python code of our analysis is here: \url{https://shorturl.at/QZ9v7}.

\section{Delayed Choice Entanglement Swapping}\label{sec:dces}

\begin{figure*}
    \centering
    \includegraphics[width=\linewidth]{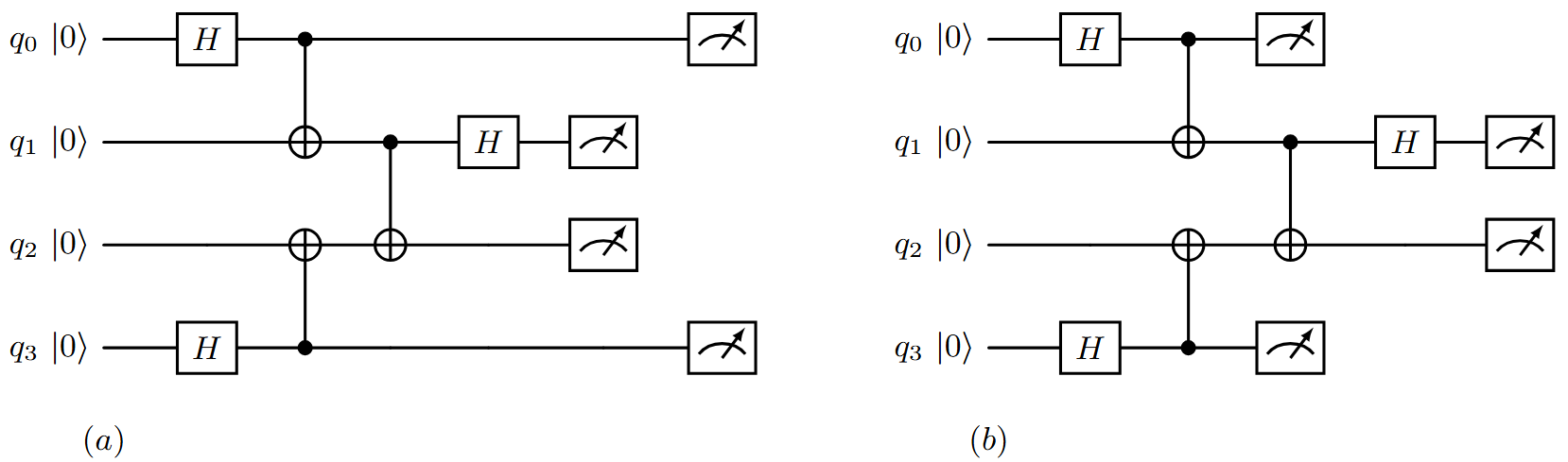}
    \caption{Circuit diagram of quantum entanglement swapping. In (a), the conventional entanglement swapping \cite{coecke2014logic} occurs, with Charlie performing the Bell measurement before Alice and Bob does. In (b), the order of measurement is reversed, with Alice and Bob measuring their qubits before Charlie. In both the circuits, all the qubits are starting at $\ket{0}$ state. These two circuits are equivalent as the two sets of measurements are parallel, non-overlapping, and independent of each other.}
    \label{fig:dces}
\end{figure*}

To describe our protocol, we start by providing a concise overview of the initial delayed-choice entanglement-swapping technique, as explained by Peres \cite{peres2000delayed}. In accordance with Peres, we examine the most basic iteration of this protocol. This protocol aims to establish entanglement and eventually generate correlated bits between two parties, Alice and Bob, who choose a third party, Charlie, to mediate the entanglement generation. Consider three individuals, namely Alice, Bob, and Charlie, as seen in the circuit diagram in Figure \ref{fig:dces}. 
Two Bell Pairs $q_0 q_1$ and $q_2 q_3$ are present such that $q_0$ is in the custody of Alice, $q_3$ with Bob, and $q_1$ and $q_2$ with Charlie. 
Qubits $q_0$ of Alice and $q_3$ of Bob get entangled when Charlie performs a joint Bell-measurement on his qubits $q_1$ and $q_2$. However, the specific relationship between Alice and Bob's qubits cannot be known until Charlie reveals the outcome of his Bell state measurement. Alice and Bob's photons are perfectly entangled, disregarding any errors or decoherence. This exact entanglement allows them to be used for various entanglement-based protocols, such as cryptography using the E91 protocol \cite{ekert1991quantum}. 

Alternatively, Charlie has the option to measure his photons separately. By adopting this approach, there would be a complete absence of entanglement between the photons belonging to Alice and Bob. Consequently, any correlations seen would be purely coincidental and not indicative of any underlying connection. If Charlie attempted to conceal his independent measurements, Alice and Bob might easily expose the deceit by using the conventional method of conducting measurements and announcing certain findings subsequent to receiving Charlie's 'correction' signals. 

Peres' idea goes beyond entanglement swapping by recognizing that Alice, Bob, and Charlie might potentially be space-like separated. This suggests that one can discover a frame of reference when the sequence of qubit measurement is modified, as shown by the Gisin group \cite{zbinden2001experimental}. In the preceding discourse, we used the assumption that Charlie conducted his measurement before Alice and Bob did. Alternatively, it is possible for Alice and Bob to conduct their measurements prior to Charlie (Figure \ref{fig:dces}(b)). In this scenario, the outcome of their measurements itself is not affected by whether or not Charlie performed either the Bell state or independent measurement, but the degree of correlation between Alice and Bob's measured bits is dependent upon the kind of measurement performed by Charlie. This serves as the foundation for the delayed-choice entanglement-swapping protocol and is crucial for the advancement of our ideas. In the delayed-choice entanglement swapping technique, Alice and Bob perform measurements on their qubits and record the outcomes using classical storage. At a later point, Charlie has a decision to either measure his qubits separately or do a Bell state measurement. The measurement outputs are either associated in a known manner or wholly uncorrelated, depending on the measurement option chosen by Charlie and the random basis choice of Alice and Bob.

\section{Setup}\label{sec:dcsetup}

\begin{figure*}
\begin{minipage}{\linewidth}
\begin{algorithm}[H]
\caption{Delayed Choice Quantum Network Protocol}\label{alg:dcqn}
    \begin{algorithmic}[1]
        \ForAll{end-users}
            \State Charlie prepares Bell pairs encoded in surface codes
            \State Charlie loads each half of the Bell pairs into separate sets of qLDPC code blocks via teleportation
            \State Charlie stores one set of qLDPC blocks and transports the other set to destination
            \State End-user puts in a purchase order for $n_{bits}$ number of bits as per their requirement
            \For{i from 1 to $n_{bits}$}
                \State qATM at the destination unloads a logical qubit from a qLDPC block onto a surface code  
                \State qATM measures the logical surface code qubit to get a bit
                \State qATM transfers the bit onto the user's smartphone
                \State Bit stored in the end-user's smartphone for later use.
            \EndFor
        \EndFor
        \State Publish measurement bases
        \For{users Alice and Bob wishing to connect}
            \State Alice and Bob decide to correlate $n_{key}$ number of bits
            \For{i from 1 to $n_{key}$}
                \State Charlie unloads two corresponding qLDPC qubits onto two surface code patches
                \State Charlie performs logical Bell measurement on surface code patches 
                \State Charlie broadcasts the measurement result to Alice and Bob, resulting in a correlated key bit
            \EndFor
            \State Alice and Bob perform standard reconciliation techniques and get the final secure key
        \EndFor
    \end{algorithmic}
\end{algorithm}
\end{minipage}
\end{figure*}

\begin{figure*}
    \centering
    \includegraphics[width=0.3\linewidth]{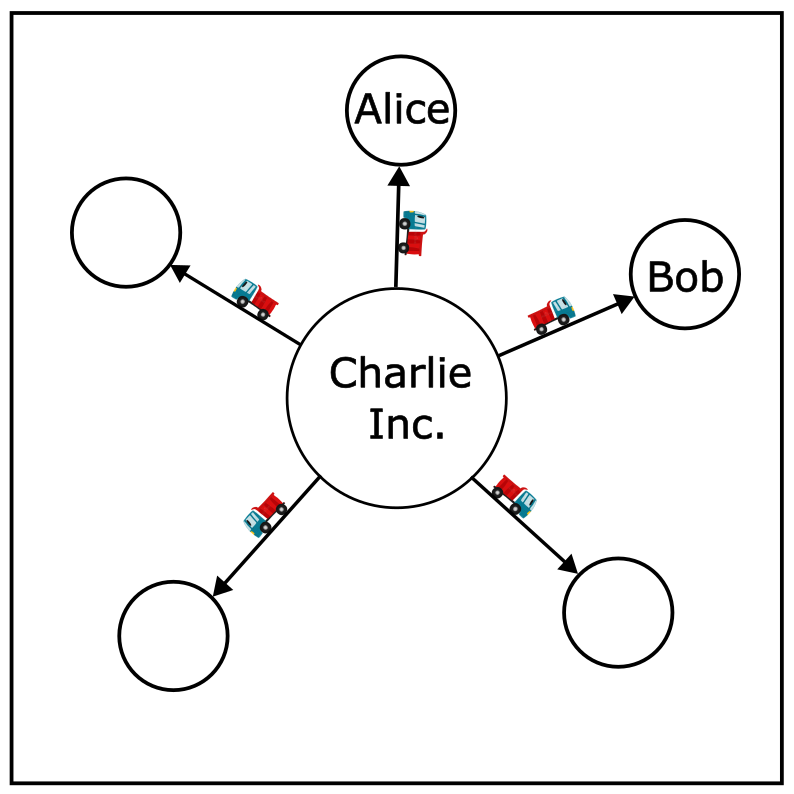}
    \includegraphics[width=\linewidth]{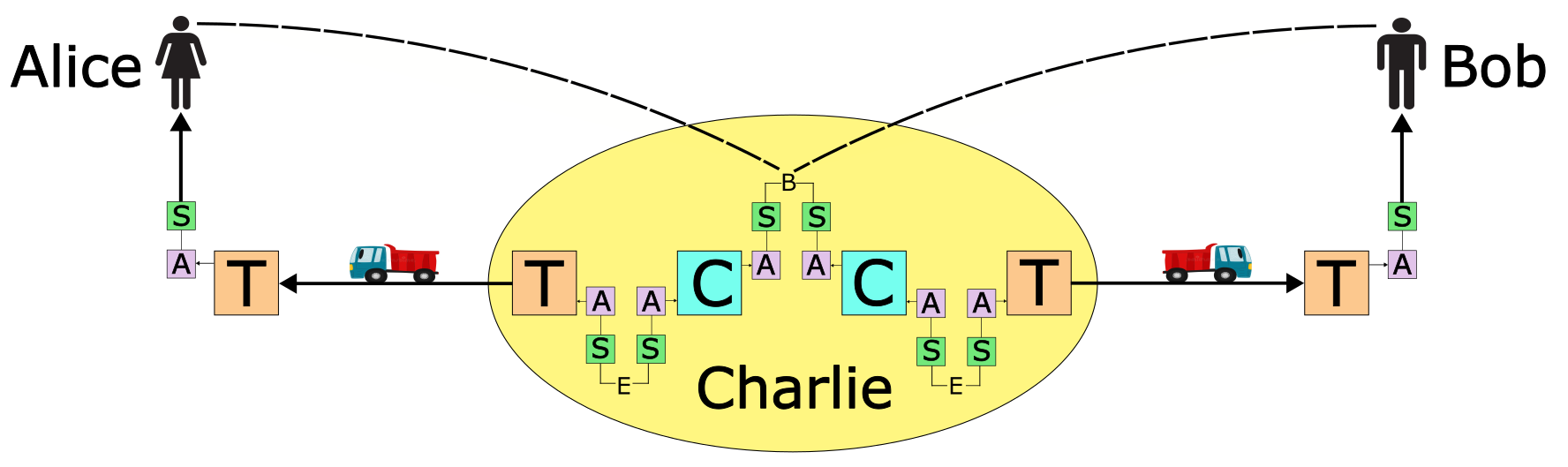}
    \caption{(Above) A schematic of a 5-node delayed choice quantum network with Charlie Inc. in the center. (Below) Delayed choice entanglement swapping protocol between two nodes Alice and Bob. S represents the surface code patches, A refers to the ancilla patches encoded in HGP code, C denotes Charlie's qLDPC memory and T represents the qLDPC memory transported to the end-nodes. Here, E represents the initial entangled Bell-pair and B denotes bell measurement.}
    \label{fig:dcqn}
\end{figure*}

We now assume that Charlie is replaced by a company Charlie Inc. whose business is to store and distribute entangled states. This requires Charlie Inc. to have large-scale long-lived quantum memory and a means of distributing entangled states. Our protocol is detailed in Figure \ref{fig:dcqn}.
and is also described concisely in pseudocode \ref{alg:dcqn}.
Charlie Inc., acting as a central node, first prepares Bell pairs in surface codes and teleports each half into separate qLDPC code blocks via a qLDPC (HGP) ancilla. This is done because direct computations and measurements for qLDPC codes are not yet well defined at the time of this writing. Charlie Inc. then keeps one set of neutral-atom encoded qLDPC blocks and sends the other set to an end-user via physical transport, like a train or a truck. This process is done for several such end-users at different destinations. This is called the "sneakernet" method. 
At the user's end, the user purchases a certain number of bits based on their needs. At the user's end, a quantum ATM (qATM) ``unloads" the incoming qLDPC blocks by sequentially teleporting the constituent qubits onto a surface code, followed by measuring the surface code randomly in X or Z bases, and commercially selling the resulting classical bits to users, which is in turn stored in a device belonging to the user, such as a smartphone. For our analysis, we assume that the process of transferring classical bits onto the end-users' classical devices is secure. The measurement bases are then made public. 

Among several end-users in the network, let's say two of them, namely Alice and Bob, want to connect at a later time. 
They decide upon how many bits they need to correlate for their requirement. Charlie Inc. then converts the corresponding logical qubits in the qLDPC blocks on his end into surface code patches and performs Bell measurements on those patches, thereby correlating Alice and Bob's bits. Finally, Alice and Bob use their correlated bits to generate a draft encryption key. Depending on their security requirement, they can enhance the key's security using CHSH tests \cite{clauser1969proposed} and privacy amplification, sacrificing some bits in the process. The remaining bits form a secure encryption key for end-to-end message encryption. This approach turns entanglement into a tradeable commodity and allows users with only classical hardware to participate in quantum cryptography protocols.  
For example, Alice measures a number of qubits, labeled ${A_1, A_2,...,A_n}$ and informs Charlie Inc. which qubit measurement results she now owns. Charlie Inc. then records the owner of the qubits and stores the other half of the Bell pairs in long-term quantum storage. Similarly, Bob buys the rights to measure qubits ${B_1,B_2,...,B_n}$ and Dave buys the rights to measure qubits ${D1,D_2,...,D_n}$. At this stage, neither participant has decided who they wish to exchange correlations with, they have simply ‘topped up’ their future quantum potential i.e. the bits with which they can perform an unconditionally secure one-time-pad with another user of their choice in the future. If Alice and Dave wish to generate entangled data for some protocol, then they need to publicly declare to Charlie Inc. how many topped up bits they want to have correlated, then Charlie Inc. performs the entanglement swap between the appropriate (stored) Bell pairs and broadcasts the measurement results. If Alice and Dave wish to check Charlie Inc.’s fidelity, they may sacrifice some of their bits, and assuming they are satisfied, can proceed with standard key distillation and privacy amplification \cite{bennett1995generalized}. If Alice then wishes to share secrets with Bob, then again, she need only declare to Charlie Inc. how many bits she wants to correlate (of course, she cannot use measurement results that have previously been used with Dave), and proceed as before.

\section{Resource Estimation}\label{sec:dcresourceestimation}

In this section, we derive closed-form expressions for logical failure rates, time scales, and logistics concerning the protocol. For our analysis, we consider city-wide and district-wide distances, hence we will be considering trucks or train cars for transport of quantum memories. We consider both trucks and train cars to be equivalent in carrying capacity, transport speeds, etc., and may tend to use the terms interchangeably. Throughout this section, we consider the processes and operations occurring in the quantum error-corrected qLDPC memories as they go through all the stages (loading, transport, and unloading) of the protocol.

For our protocol, we select the quantum LDPC codes constructed from the hypergraph product of classical codes based on (3, 4)-regular bipartite graphs with strong expansion properties \cite{xu2024constant}. These graphs are selected randomly, and the resulting quantum codes exhibit a minimum encoding rate of 1/25. In ref. \cite{xu2024constant}, the authors discuss the implementation of computation with qLDPC codes using neutral atoms, which provide an order of magnitude improvement in the number of physical qubits used compared to surface codes. They introduce a technique by which logical qubits can be teleported between the surface codes and qLDPC code blocks, thus aiding in the possibility of computation and individual logical qubit measurements, which becomes essential for the implementation for our protocol. The authors also provide closed-form expressions for various parameters such as logical failure rates and cycle times, which we incorporate in our resource analysis.

\textbf{Rearrangement of atoms:} For a $\big[[n,k,d]\big]$ hypergraph product (HGP) code with a coding rate $r = k/n$ where $n$ is the number of physical qubits, $k$ is the number of logical qubits, and $d$ is the code distance, each syndrome extraction (SE) cycle takes time $8t_r$ \cite{xu2024constant}, where the factor 8 represents the number of atomic rearrangements needed to complete one error correction cycle, and $t_r$ is the time required to move the atomic qubits around for one rearrangement \cite{xu2024constant}. By overlapping the syndrome measurements with the atom repositioning of the next iteration, we can avoid extending the total execution time. We express $t_r$ as a function of $n$, given by:
\begin{equation}
t_r(n) \approx 2\tau_t \log L + (3+2\sqrt{2})\sqrt{\frac{6Ld_p}{a_p}}
\end{equation}
where $\tau_t = 50\mu s$ is the transfer time between the atomic traps, $a_p = 0.02 \mu m \mu s^{-2}$ is the maximum acceleration rate of the atom, $d_p = 5 \mu m$ is the spacing between the atoms in the qLDPC grid, and $L \approx \sqrt{n}$ is the number of atoms in a line to be rearranged.
The gate times of neutral atom architectures are in the order of microseconds \cite{wintersperger2023neutral} while $t_r$ is in the order of several milliseconds. Therefore $t_r$ becomes the dominant rate-limiting time scale and the gate times can be ignored while computing neutral atom qLDPC cycle times.

\textbf{Logical Failure Rates (LFRs):}The Logical Failure Rate (LFR) $R_L$ per error correction cycle for a [$n$, $k=n/25$, $d=\sqrt{k}$] HGP code is given by \cite{xu2024constant}:
\begin{align}
    R_L(n) = 0.07 (p_g / 0.006)^{0.47n^{0.27}}
\end{align}
where $p_g$ is the physical gate error and $n$ is the number of physical qubits, $k=n/25$ is the number of logical qubits, and $d=\sqrt{k}$ is the code distance. Ref\cite{xu2024constant} considers HGP codes such that $k=n/25$ and $d=\sqrt{k}=\sqrt{n}/5$. We consider this configuration as well in our analysis.
Similarly, the Logical Failure Rate (LFR) $R_S$ per error correction cycle for a surface code of distance $d_{sc}$ is given by \cite{devitt2016high}:
\begin{align}
    R_S = 0.3(70 p_g)^{\frac{d_{sc}+1}{2}}
\end{align}

Let $n_a$ be the number of physical qubits in the ancilla HGP code. Let $n_t$ be the number of physical qubits in the qLDPC memory to be transported. Let $n_c$ be the number of physical qubits in the qLDPC memory of Charlie Inc.. We make two key assumptions in our analysis. First, we consider the number of physical qubits in Charlie Inc.'s qLDPC memory ($n_c$) and the transported qLDPC memory ($n_t$) to be equal, denoting both as $n_m$, where 'm' stands for memory. Secondly, we consider the distance of the computational surface code ($d_{sc}$) as a function of the qLDPC memory size, specifically $d_{sc} = \sqrt{n_m}/5$, as used by the analysis in \cite{xu2024constant}.

Throughout our derivations, we use the following binomial approximation:
\begin{align}
    (1-p)&=(1-p_1)^{n_1} (1-p_2)^{n_2}... \nonumber\\ => p & \approx n_1 p_1 + n_2 p_2 + ... \nonumber\\
    & \forall \, 0<p_i<<1
\end{align}
This approximation is valid due to the logical failure rates being much lower than 1, due to which logical errors can be propagated via simple addition.

\textbf{Initialization of qLDPC code patches:}
 First, Charlie Inc. needs to initialize two qLDPC patches, one ($C$) for his memory and the other ($T$) for transporting to the destination node. The Logical Failure Rate (LFR) $R_0$ for this is given by: 
 \begin{align}
     R_0 = 2 \frac{\sqrt{n_m}}{5}R_L(n_m) 
 \end{align}
Here $d=\sqrt{n_m}/5$ is the code distance, which is also the number of error-correction cycles, and $R_L(n_m)$ is the LFR per error correction cycle. We assume that these memories are initialized in parallel with the previous memory batch being loaded onto the truck. Therefore, we do not count it into our time scales.

\textbf{Bell Pair Creation:}
Next step, we need to create a Bell-pair. Now, For creating a single Bell Pair, we need two surface codes and $2\frac{\sqrt{n_m}}{5}$ error-correction cycles. The Logical Failure Rate (LFR) $R_1$ for this is given by \cite{leone2023upper,devitt2016high}:
\begin{align}
    R_1=2\frac{\sqrt{n_m}}{5} R_S 
\end{align}
Since loading the current bell pair and creating the next bell pair are done simultaneously, we don't count this time scale.

\textbf{Loading of the Bell Pair:} 
For the loading of both halves of the surface-coded Bell-Pair onto their respective qLDPC patches $C$ and $T$, we need to teleport it from the surface code to the HGP patch via an ancilla HGP. The ancilla HGP is of the configuration $\big[[n_a=n_m/25, k=1, d=\sqrt{n_a}]\big]$. For the teleportation, we need first to initialize the ancillae. then we need to do two lattice surgery procedures, one merge-and-split between the surface code and the ancilla HGP patch, and another between the ancilla HGP and the qLDPC patches. the LFR for loading onto the transport qLDPC is given by:
\begin{align}\label{eqn:R2}
    R_{2} &= 2 \big(\sqrt{n_a}R_L(n_a) \nonumber\\ &+  \frac{\sqrt{n_m}}{5} (R_S+R_L(n_a)) \nonumber\\ &+ \frac{\sqrt{n_m}}{5} (R_L(n_a)+\frac{n_m}{25}R_L(n_m))\big)
\end{align}
where the first term is for the ancilla initialization involving $d=\sqrt{n_a}$ rounds of error correction, the second term is for lattice surgery between surface code and ancilla involving $d=\frac{\sqrt{n_m}}{5}$ error correction cycles, with LFR per cycle for the surface code being $R_S$ and that of the ancilla HGP being $R_L(n_a)$, and the third term is for that of ancilla and the memory patch. The patch sizes are chosen in such a way that all three patches (memory, ancilla, and the surface code) have the same code distance, and hence require the same number of error correction cycles. Here the $\frac{n_m}{25}$ factor appears since the first loaded qubit will undergo these sets of loading QEC cycles $\frac{n_m}{25}$ times, which propagates its error. The term 2 arises because we need to teleport two halves of the Bell-pair, one to the transport memory $T$ and the other one to Charlie Inc.'s memory $C$.
The time $T_2$ required for this is given by:
\begin{align}
    T_2 &= \frac{n_m}{25}\bigg(  \sqrt{n_a}t_{cyc}(n_a) \nonumber\\ &+ \frac{\sqrt{n_m}}{5} t_{cyc}(n_a) + \frac{\sqrt{n_m}}{5} t_{cyc}(n_m) \bigg)
\end{align} 
where $t_{cyc}=8t_r$ is the time taken to complete one error correction cycle. Here we combine parallel cycles, such as that happening simultaneously on surface code and ancilla, and on the ancilla and the qLDPC memory.

\textbf{Storage and Transport:} 
Let's say the destination node is $T_3$ units of time drive from Charlie Inc.. Throughout the transit, $M_t = \frac{T_3}{T_{cyc}(n_m)}$ number of error correction cycles happen on the transport memory $T$. The LFR $R_3$ occurring during the transport is given by:
\begin{align}
    R_3 = \frac{T_3}{T_{cyc}(n_m)} R_L(n_m)
\end{align}
Here, the idling errors, which emerge as an adjustment to the value of $p_g$, have been incorporated into $R_L$, and that a scenario has been considered where it is beneficial to perform QEC cycles as often as
possible. This is explained in section \ref{sec:results} under the topic of idling errors (Eq.\ref{eqn:idling_errors}).

\textbf{Unload and Measure at qATM:}
Once the transport memory arrives at a qATM in the destination node, we need to initialize the surface code and the ancilla, and then sequentially teleport the logical qubits from the qLDPC to the surface code via the ancilla, and measure the surface code qubit. The LFR for this is given by:
\begin{align}
    R_4 &= \frac{\sqrt{n_m}}{5} R_S \nonumber\\ &+ \sqrt{n_a}R_L(n_a) \nonumber\\ &+  \frac{\sqrt{n_m}}{5} (R_S+R_L(n_a)) \nonumber\\ &+ \frac{\sqrt{n_m}}{5} (R_L(n_a)+\frac{n_m}{25}R_L(n_m))   + R_S
\end{align}
This is similar to Eq. \ref{eqn:R2}, with an additional term $R_S$ in the end which is the 1-round LFR that occurs due to measurement. 
The time taken for this is:
\begin{align}
    T_4 &= \frac{n_m}{25}\bigg( 6 \frac{\sqrt{n_m}}{5} t_g \nonumber\\ &+ \sqrt{n_a}t_{cyc}(n_a) \nonumber\\ &+ \frac{\sqrt{n_m}}{5} t_{cyc}(n_a) \nonumber\\ &+ \frac{\sqrt{n_m}}{5} t_{cyc}(n_m) + 6 t_g \bigg) 
\end{align}
Measuring a surface-code logical qubit involves one round of error correction. The depth of a surface-code syndrome extraction is 6 and the physical gate time of the architecture is given by $t_g$. Hence the last term.

Now, the total network LFR $R_{net}$ of the processes so far, and the total time $T_{tot}$ taken for generating this correlation is:
\begin{align}
    R_{net} &= \sum_{j=0}^{4}R_j, \nonumber\\ T_{tot} &= \sum_{j=2}^{4}T_j
\end{align}

\textbf{Unload and Bell-measure at Charlie Inc.'s end:}
Now we have two such half-measured Bell-Pairs distributed between the network pairs Charlie Inc.-Alice and Charlie Inc.-Bob. So we have two memory drives, one each for each network pair. Let's say at a later time $T_{later}$, Alice and Bob decide to correlate their bits.
For calculation purposes, we assume that this decision has been made by the time $T_{later}=T_{tot}$. We assume that by this time, the end users Alice and Bob have decided to communicate with Charlie Inc. and have informed him accordingly. Let the total number of error correction cycles $M_{tot_{c}}$ be done by Charlie Inc.'s memory drives during this time. This is given by $M_{tot_{c}} = T_{later}/t_{cyc}(n_m)$. Since we have two drives, the LFR $R_{store}$ is given by:
\begin{align}
    R_{store}= 2 M_{tot_{c}} R_L(n_m) = 2 \frac{T_{tot}}{t_{cyc}(n_m)} R_L (n_m)
\end{align}
The factor of 2 occurs because there are two memory drives.
Now the qubits need to be teleported onto surface code via an ancilla. The LFR $R_{unload}$ of that is given by:
\begin{align}
    R_{unload} &= 2 \bigg( \frac{\sqrt{n_m}}{5} R_S \nonumber\\ &+ \sqrt{n_a}R_L(n_a) \nonumber\\ &+ \frac{\sqrt{n_m}}{5} (R_L(n_a) + \frac{n_m}{25}R_L(n_m)) \nonumber\\ &+ \frac{\sqrt{n_m}}{5} (R_L(n_a)+R_S) \bigg)
\end{align}
Now for Bell-Measurement, we require $4\frac{\sqrt{n_m}}{5}$ QEC cycles \cite{horsman2012surface}. The LFR $R_{BM}$ for this is given by:
\begin{align}
    R_{BM}=4 \frac{\sqrt{n_m}}{5} R_S
\end{align}
This generates a correlation between Alice and Bob.

\textbf{Total LFR:} So, therefore, the LFR $R_{tot}$ of the overall final correlation between end users Alice and Bob is given by:
\begin{align}
    R_{tot}=R_{BM} + R_{unload} + R_{store} + 2 R_{net}
\end{align}
And the success probability $F_{final}$ of this correlation is given by:
\begin{align}
    F_{final} = 1 - R_{tot}
\end{align}

\textbf{Logistics:}
Logical Bit-rate $r_L$ per qLDPC block is given by:
\begin{align}
    r_L = \frac{n_m}{25T_4}
\end{align}
This is the number of logical qubits from the qLDPC block unloaded and measured per unit time. It takes $T_4$ amount of time to unload one entire qLDPC patch consisting of $\frac{n_m}{25}$ logical qubits.

Let $r_e$ be the desired target entangled bit (E-Bit) rate. Then, the number $N$ of parallel qLDPC blocks required to achieve $r_e$ is given by $N=r_e/r_L$. 
The blocks get depleted every $T_4$ units of time. Therefore, we need to replace $N$ blocks every $T_4$ units of time. Let's say a truck or a train car has the carrying capacity of $B$ blocks, where $B=n_t/n_m$ where $n_t$ is the total number of physical qubits the truck or train car can carry. We need $N/B$ trucks to be released by Charlie Inc. every $T_4$ units of time, for one qATM destination. if we are assuming $S$ number of qATM destinations, then the number of trucks $n_{truck}$ released by Charlie Inc. every $T_4$ units of time is given by $n_{truck}=SN/B$. 

The life-cycle of a truck involves $T_2$ units of time to load the qLDPC patches, $T_3$ time to transit, $T_4$ time to unload and get the logical qubits measured, and $T_3$ again to move back to Charlie Inc.'s hub for the next round. So, the total number of trucks $N_{truck_{tot}}$ required for the whole network to run continuously is given by:
\begin{align}\label{eqn:ntrucktot}
    N_{truck_{tot}} = \frac{SN}{(n_t/n_m)} \frac{T_2+ 2 T_3 + T_4}{T_4}
\end{align}

\section{Results and Discussion}\label{sec:results}

The required E-bit rate ($r_{e}$), or bandwidth, is a critical parameter in our system design. Contemporary communication standards provide context for our bandwidth requirements. For instance, voice communications necessitate 87 kbps, standard definition video conferencing requires 128 kbps, and high-definition video conferencing demands up to 4 Mbps \cite{voice_comms}. In the realm of text messaging, an average SMS of 120-160 characters consumes approximately 16,000 bits \cite{sms}, which we consider for our analysis. Note that this is somewhat contrived, but it gives us an anchor to do our analysis. To determine $r_{e}$, we consider several factors: the frequency of one-time pad recharging, the practical time a user might wait at a quantum ATM while recharging ($t_{wait}$), and the average number of bits needed between refills ($b_{req}$). The number of end-users at a destination ($n_{users}$) is constrained by the number of available time slots in a day, calculated as 86,400 seconds divided by $t_{wait}$. Consequently, we express $r_{e}$ as $b_{req}/t_{wait}$.

We estimate $r_{e}$ for SMS usage as an illustrative example. Assuming an average person sends 85 texts daily \cite{avgsmsperday}, each requiring 16,000 bits, we calculate $b_{req}$ as $85 \times 16,000 = 1,360,000$ bits. With a $t_{wait}$ of 10 minutes (600 seconds) every day, $r_{e}$ equals 1,360,000/600 = 2.3 kbps or 2.3 KHz.

The gate error rate $p_g$ for neutral atom architectures varies from $0.001$ to $0.005$. Quantinuum, a quantum computing company, has demonstrated a one-qubit gate fidelity of $0.999979$ and a two-qubit gate fidelity of $0.99914$ \cite{easttom2024trapped}.  Therefore it would be reasonable to consider $p_g=1 - 0.9992 = 0.0008$ as a near-term realizable optimistic estimate.

\textbf{Idling Errors:} Idling errors can be modeled by adjusting the two-qubit gate error rate $p_g$ as follows \cite{xu2024constant}:
\begin{align}\label{eqn:idling_errors}
    p_g \rightarrow p_g \left(1 + \frac{3 t_r(n)}{0.005 T_c} \right)
\end{align}
where $T_c = 10$s is the coherence time of neutral atom qubits. If we plug in a worst-case scenario of a patch size of $100,000$, we see that $p_g$ gets scaled up by a factor of $1.2754$ i.e. by $27.54\%$. Considering our original estimate of $p_g = 0.0008$, the new $p_g$ becomes $p_g = 0.0008*1.2754 = 0.001$, which we use for our analysis.

Unlike most other architectures, neutral atoms have significantly slower 1-qubit gates than 2-qubit gates. Hence we take the 1-qubit gate time $t_g=2 \mu s$ \cite{wintersperger2023neutral}. Considering the form factor of Pasqal's processor \cite{pasqal} and their claim of producing a $10^4$ qubit unit by 2026, we take a rough estimate for our calculations that any truck or train car in the near future can carry a total of $n_t=1,000,000$ physical qubits. We also set the number of qATM destinations $S=5$. In this article, we assume intra-metropolitan distances, with typical transport times ranging from one to three hours. We take the teleportation ancilla HGP patch size $n_a=\frac{n_m}{25}$ because it maintains the same distance as the qLDPC code and the surface code, and its edge is as wide as the size of the logical operators in the qLDPC code \cite{xu2024constant}.

\begin{figure*}
    \centering
    \includegraphics[width=\linewidth]{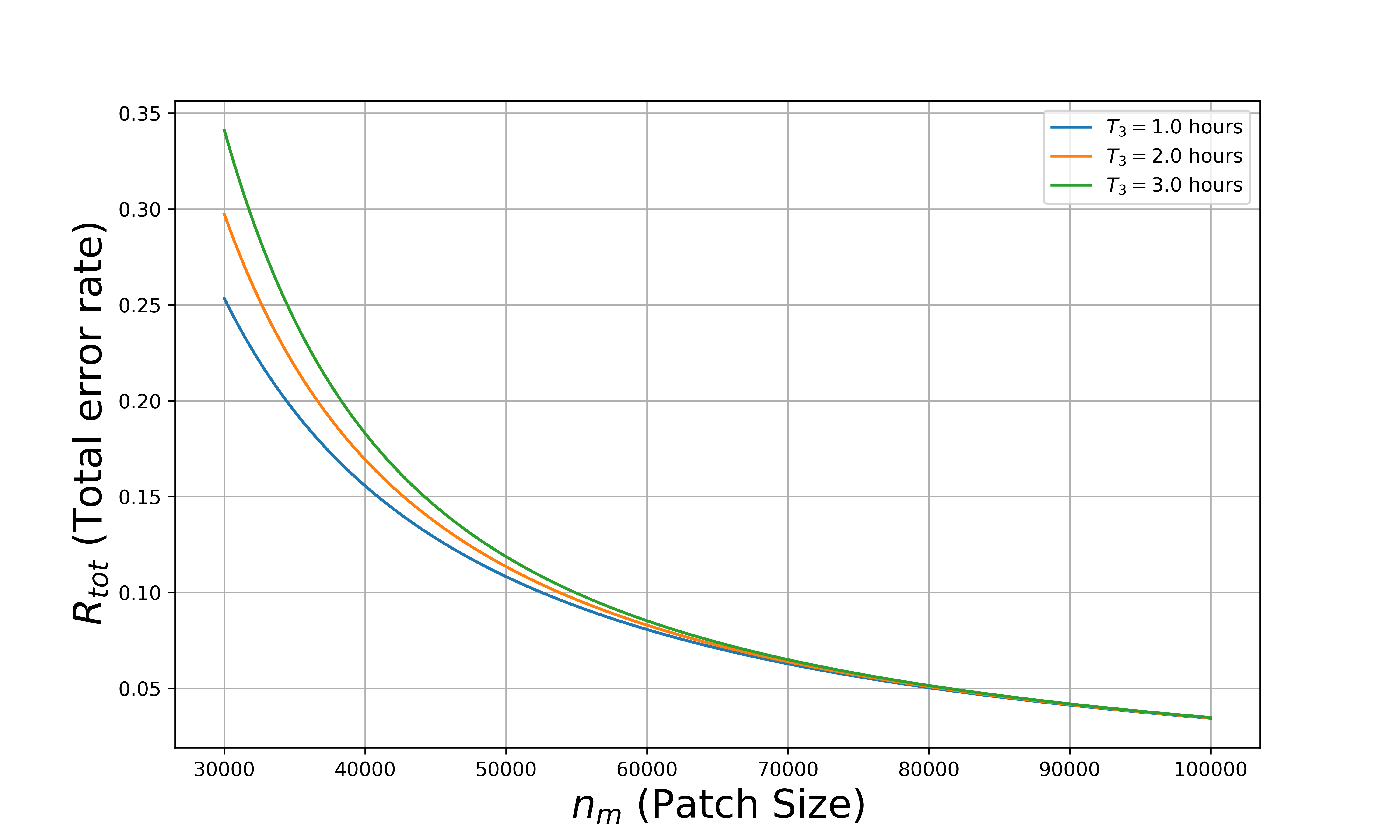}
    \caption{The overall Logical Error Rate $R_{tot}$ plotted against qLDPC patch size $n_m$ for various transport times $T_3$, showing an inverse relationship between the total logical error rate $R_{tot}$ and the size of qLDPC patches. To obtain fidelity levels exceeding 80\% (corresponding to $R_{tot}<0.2$), patch sizes of approximately 35,000 qubits or more are necessary. Further improvements in fidelity can be achieved with larger patches: those containing over 53,000 qubits can reach fidelities surpassing 90\%, while patches comprising 80,000 qubits or more are capable of attaining fidelity levels above 95\%.}
    \label{fig:rtot_vs_nm}
\end{figure*}

From Figure \ref{fig:rtot_vs_nm}, 
we can see that to achieve a lower total logical error rate $R_{tot}$, we need larger qLDPC patches. Larger patches have greater code-distance, hence higher tolerance to errors. To achieve greater than 80\% fidelity (i.e. $R_{tot}<0.2$), we need patch sizes of at least about 35,000 physical qubits (1400 logical qubits). Patch sizes over 53,000 physical qubits (2120 logical qubits) can achieve fidelities of over 90\% while 80,000-qubit-sized patches (3200 logical qubits) can give greater than 95\% fidelity. Higher the fidelity, lower is the number of bits wasted for standard reconciliation and privacy amplification, and with fidelities over 95\%, this number becomes negligible \cite{bennett1992experimental}. Pasqal, an enterprise dealing with neutral atom quantum architectures, claims that devices with a capacity of 10,000 physical qubits will be a reality in the next couple of years \cite{pasqal}, so it would be reasonable to assume larger patch sizes of the order of high $10^4$ physical qubits to be achievable in the next few years.

\begin{figure*}
    \centering
    \includegraphics[width=\linewidth]{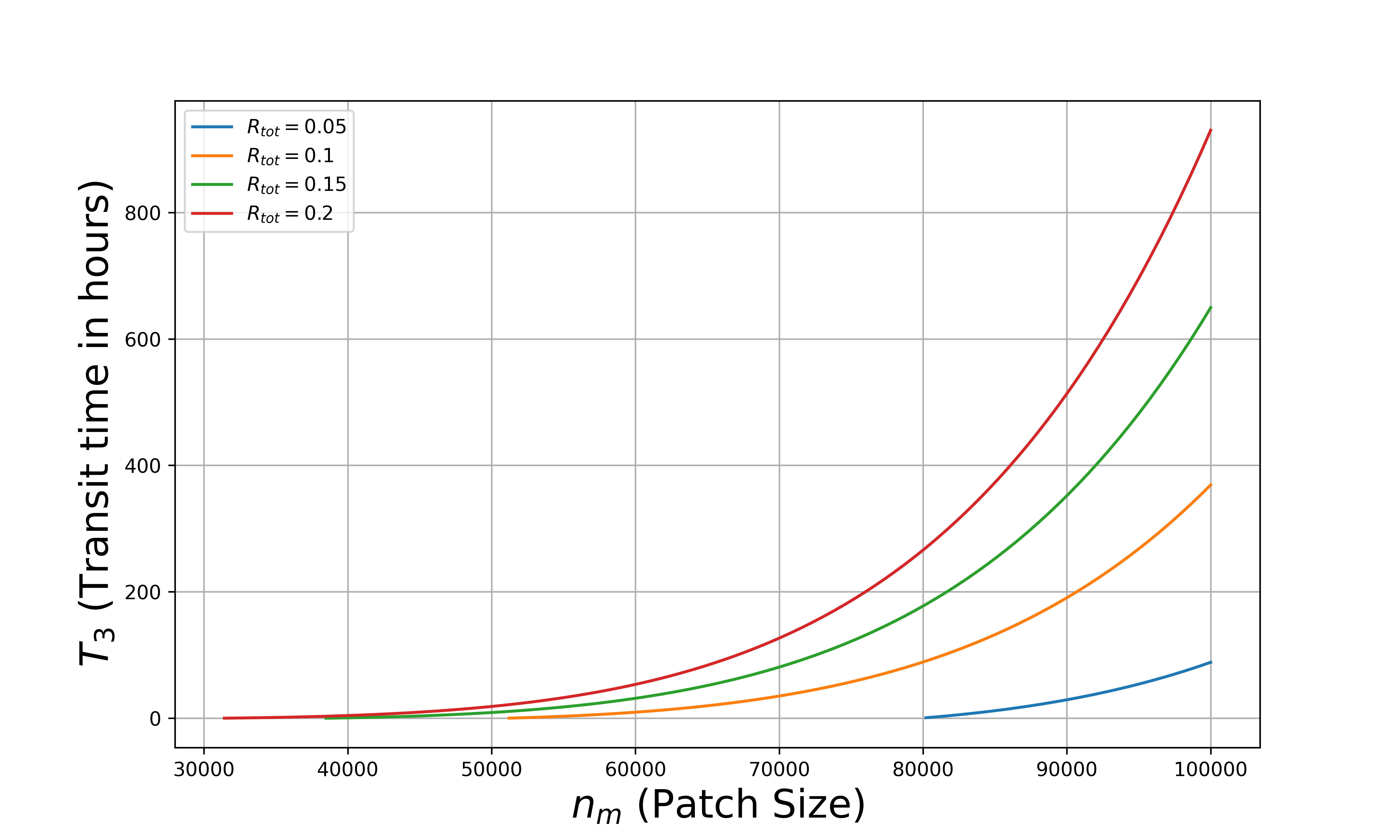}
    \caption{Transport time $T_3$ plotted against qLDPC patch size $n_m$, for various tolerable LFRs. The analysis demonstrates a positive correlation between patch size and the maximum tolerable transport time $T_3$. This relationship stems from the enhanced logical error tolerance of larger patches, which enables longer-distance transport. The plots exhibit a lower limit on patch size, at which the $T_3$ value becomes zero. This occurs when the sum of all error terms except $R_3$ exceeds the target error rate, regardless of $T_3$. In such cases, the numerical solver attempts to compensate by assigning negative $T_3$ values, indicating that the target $R_{tot}$ is unattainable for patches smaller than this threshold.}
    \label{fig:t3_vs_nm}
\end{figure*}

From Figure \ref{fig:t3_vs_nm}, 
we can observe that the protocol can tolerate longer transport times $T_3$ with larger patch sizes. This is because larger patch sizes have greater logical error tolerance, and hence can be transported to longer distances. It can also be observed that the plot truncates below a certain patch size when the value of $T_3$ comes down to zero. This is because, for any $n_m$ below that patch size, all terms in $R_{tot}$ except the $R_3$ term push the $R_{tot}$ greater than the target error rate. Due to this, the solver function in our plot code plugs in a negative $T_3$ to pull down the $R_{tot}$ to match the target. This means that the target $R_{tot}$ is not achievable for patch sizes below this cutoff size. From both protocols, we see that the error rate $R_{tot}$ increases with an increase in transport time due to a greater number of error correction cycles.

\begin{table}
\centering
\begin{tabular}{ | c | c | c | } 
  \hline
  \textbf{QEC Code}  & \textbf{Number of Trucks} & \textbf{Cost Per Bit}\\ 
  \hline
  qLDPC & 7444 & USD \$1.40 \\
  \hline 
  Surface Code & 31920 & USD \$5.99 \\
  \hline
\end{tabular}
\caption{\label{tab:logistics} Table showing the number of trucks required and the cost of each bit sold rounded to the nearest cent, for qLDPC-based protocol and a  pure Surface-Code-based protocol. Here, we have taken the transit time $T_3$ to be 90 minutes, network bandwidth to be 2.3 kHz, hourly rent per truck as USD \$150, and the total number of end-nodes to be 5. The cost is calculated as per Eq. \ref{eqn:cost}.}
\end{table}

\begin{figure*}
    \centering
    \includegraphics[width=\linewidth]{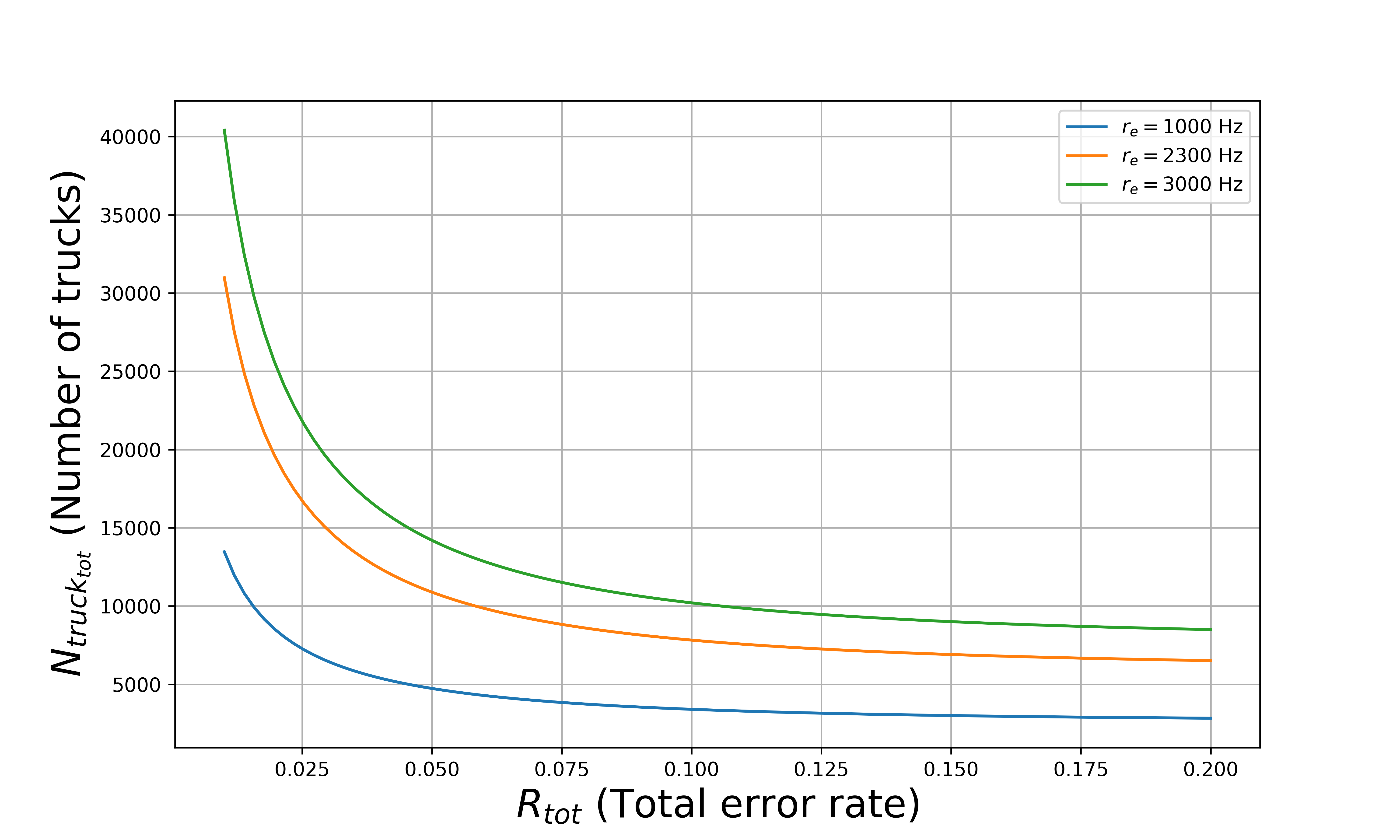}
    \caption{Number $N_{truck_{tot}}$ of trucks or train cars required for the network, as a function of overall LFR $R_{tot}$, showing an inverse relationship between the network's tolerable error rate $R_{tot}$ and the required number of transport vehicles (trucks or train cars) for a fixed transit duration of 90 minutes. As the acceptable $R_{tot}$ increases, the fidelity requirement decreases, and smaller patch sizes become viable. These reduced patch sizes correspond to lower code distances, necessitating fewer error correction rounds. Consequently, the unloading process becomes faster, reducing the number of parallel qLDPC patches needed. This cascade effect ultimately results in a decreased demand for transport vehicles to maintain the desired bandwidth.}
    \label{fig:trucks}
\end{figure*}

From Figure \ref{fig:trucks},
we can see that for a given transport time (90 minutes here), the number of trucks or train cars required reduces with an increase in the tolerable error rate $R_{tot}$ of the network. This is because the greater the tolerable error rate $R_{tot}$, the smaller the patch size. Smaller patch sizes are of lower code distances and hence require fewer rounds of error correction. This translates to faster unloading times, requiring a lesser number of patches in parallel, and ultimately requiring a lower number of trucks. For fidelities below 95\% (i.e. $R_{tot}>0.05$), we can run this network with under 10,000 trucks for 2300 Hz bandwidth. This is comparable to the total number of buses or train cars in an average city public transport.

\begin{figure*}
    \centering
    \includegraphics[width=\linewidth]{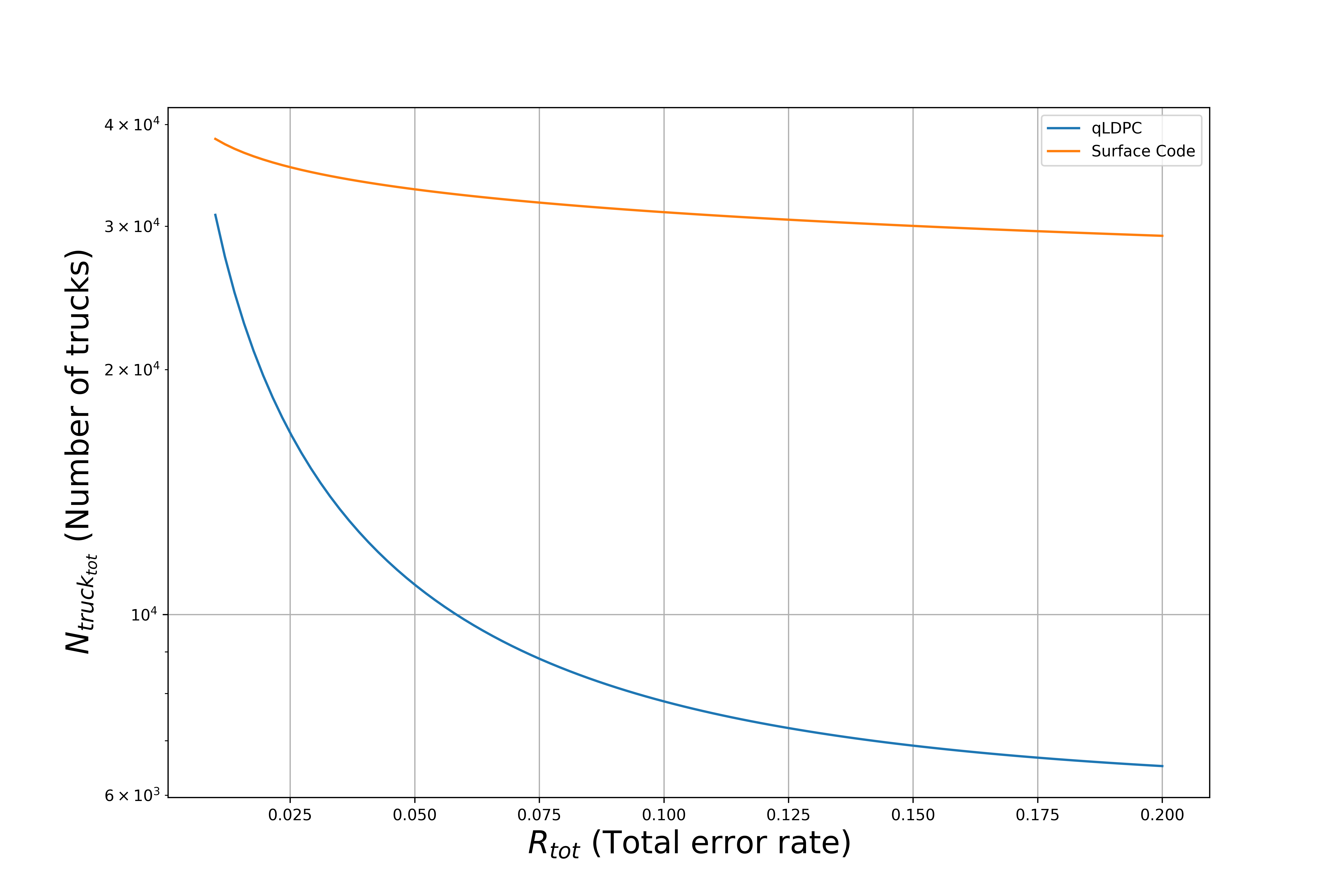}
    \caption{Resource Requirements (Number of vehicles) for Surface code vs qLDPC. The qLDPC network protocol demonstrates significantly higher efficiency in transport resource utilization compared to the surface code protocol, nearly by an order of magnitude.}
    \label{fig:sc_vs_ldpc}
\end{figure*}

Figure \ref{fig:sc_vs_ldpc}
compares the qLDPC network with the corresponding surface-code network, in terms of number of trucks, and additionally, table \ref{tab:logistics} also lists the operational unit economics $C_{o}$ in terms of the cost per bit, which is a combination of transportation cost $C_t$ and the cost $C_q$ of maintaining the quantum infrastructure i.e.:
\begin{align}\label{eqn:cost}
    C_o = C_t + C_q
\end{align}
The transport cost $C_t$ calculated as follows:
\begin{align}
    C_t = \frac{R_h N_{truck_{tot}}}{3600 r_e S} 
\end{align}
where $R_h$ is the hourly rent of one truck, which we assume to be USD \$150, $N_{truck_{tot}}$ is as described in Eq. \ref{eqn:ntrucktot}, $r_e$ is the E-bit rate or the bandwidth of the network, and $S$ is the number of network nodes or end-stations. 
Next, the total cost $C_q$ of maintaining the quantum infrastructure is given by:
\begin{align} \label{eqn:cq}
    C_q &= \frac{2 \frac{n_t}{n_m} N_{truck_{tot}} C_m}{r_e S \times 86400 \times 365} \nonumber\\
\end{align}
where $C_m$ is the yearly cost of maintaining a single qLDPC memory patch containing $n_m$ physical qubits. The numerator represents the yearly cost of maintaining all the qLDPC memory patches in the network, where the factor 2 arises from the fact that we are dealing with Bell-pairs, and $n_t/n_m$ is the number of quantum computer units on a truck. The denominator represents the number of bits emenated per year by the network. We consider an estimate of $C_m=\$ 2,000,000$ per year per quantum device unit \cite{qubitcosts}. For surface codes, the quantity $n_m$ would represent the number of physical qubits in a single quantum device, which would contain multiple surface code patches. We keep $n_m$ same for both qLDPC and surface codes.

The number of transport vehicles required by the qLDPC network is one order of magnitude lower than the number of vehicles required by the surface code protocol, and so is the cost. For example, for a minimum fidelity of $0.92$, patch size $n_m=60,000$, and a transit time of 90 minutes, the qLDPC network requires 7444 vehicles to run and costs $C_o=$USD \$1.40 per bit, whereas an equivalent surface code requires 31920 vehicles and costs $C_o=$USD \$5.99 per bit, about 4 times more. This is due to the constant encoding rates of qLDPC codes regardless of the tolerable logical error rate, unlike surface codes whose encoding rate worsens with the required tolerable logical error rate.

Nevertheless, even with qLDPC codes, the cost of \$1.40 per qubit would mean an average of \$22,400 per text message at 16,000 bits per message, which is extremely expensive, thus making the protocol in its current form commercially infeasible. High number of trucks and consequently high costs per qubit are both mainly attributable to the fact that neutral atom architectures have their physical qubits sparsely spaced in the lattice, and their gate execution times are relatively slow. If we can implement qLDPC codes on faster, more condensed architectures in the future, such as the silicon quantum dot qubits for example \cite{eriksson2013semiconductor}, we can potentially reduce the number of trucks and consequently the cost per bit by about 3 orders of magnitude. 

\section{Conclusion}\label{sec:conclusion}

Our analysis shows that in the near term, it is possible to build a multi-node Delayed-choice network, all separated by intra-metropolitan distances (with transport times of one to three hours), with a central Charlie Inc. who uses $O(10^3)$ vehicles to transport and sell entanglement-based correlations to end-users with classical hardware (such as a smartphone) at the end nodes. Assuming that Charlie Inc. has the requisite quantum memory and storage, and it is possible to distribute entanglement at a rate faster than the entanglement consumption, then many end-users can top up their results on classical storage devices, and use them at their convenience, topping up as required. This commodifies entanglement, paving the way to large-scale commercial quantum networks catering to users with non-quantum hardware. The qLDPC codes, with their compact constant encoding rates, play a major role in establishing this feasibility. 

In our analysis, we adopt certain conventions and assumptions based on recent research in quantum error correction codes. The relationship between the number of physical qubits in Low-Density Parity-Check (qLDPC) memory and Surface Code is a key consideration. We assume that a qLDPC memory with $n_m$ physical qubits is connected to a surface code with $n_m/25$ physical qubits for teleportation. This ratio is consistent with the simulations presented in ref. \cite{xu2024constant}, although it is not explicitly stated as a requirement. While it is theoretically possible to use surface codes of various sizes, we adhere to this ratio for consistency with existing literature.

Regarding the teleportation scheme for transferring logical qubits between qLDPC and surface codes, our analysis focuses on Hypergraph Product (HGP) codes. Ref\cite{xu2024constant} specifically defines this teleportation scheme for HGP codes, but not for Lifted Product (LP) codes. While it may be possible to extend this scheme to LP codes, such an extension would require additional assumptions about the ability to generate logical X operators at the edge of the patch. To maintain rigor and avoid unsubstantiated claims, we restrict our analysis to HGP codes, for which the teleportation scheme is well-defined. This decision ensures that our results are based on established protocols and avoids potential inaccuracies that could arise from extrapolating beyond the current state of knowledge in the field. Designing a teleportation protocol for these other non-HGP qLDPC code families can reduce patch sizes by a factor of two to four, reducing our resource requirements.

Another caveat is that of classically transferring measurement results onto the users' smartphones. In future research, if a protocol can be devised to teleport the logical qubits of a quantum error-correcting code patch onto a single photon, then photon measurement device can be built into smartphones, which makes the transfer quantum-secure. An alternate way, as outlined in the pseudocode \ref{alg:dcqn2} is to perform a heralded qubit transfer from each atomic physical qubit on the surface code to a photon \cite{kalb2015heralded,covey2023quantum} and measure those photons on the end-user's smartphone \cite{lowndes2021low} in a pre-defined sequence and decode to extract the logical surface-code measurement, but this method has only 0.88 fidelity and 0.69 efficiency \cite{kalb2015heralded}, which when propagated over all the atoms of a single surface code, would lead to high failure rates and high resource overheads. Therefore, development of a more efficient method of performing atom-to-photon qubit transfer would be beneficial towards a highly secure network with minimal resource overhead. 

With this kind of network, one can easily imagine creating new symmetric keys between multiple users, for example, a consumer who wishes to secretly share her banking details with a large number of different vendors, all via guaranteed entanglement-generated quantum keys. We have so far considered a single Charlie Inc., with one long-term mass storage and qATMs distributed in the field. However, as long as Charlie Inc. has set up pre-existing and refreshed entanglement, there is no reason why it is not possible for Charlie Inc. to have multiple long-term quantum memory sites in geographically distinct locations, which we term quantum hubs. With a constant refresh of entanglement between Charlie Inc.’s hubs, in principle, global scale links could be achieved with additional entanglement-swapping operations.

\bibliography{refs}

\appendix

\section{A surface-code-based delayed choice network}

For comparison, we design a surface-code-based delayed-choice quantum network. Any variable not explained here has been adopted from the main paper.

In this network, there is no teleportation involved between a separate storage code and an input code since we are dealing with surface codes directly. Hence, the storage-and-transport code is the same as the input/output code.

Let $n_{ms}$ be the number of qubits in the surface code patch. Its distance will be $\sqrt{n_{ms}}$.
Surface code LFR $R_{ss}$ per cycle is given by:
\begin{align}
    R_{ss}=0.3(70 p_g)^{\frac{\sqrt{n_{ms}}+1}{2}}
\end{align}

\textbf{Bell Pair Creation:} First, Charlie Inc. creates Bell-pairs. This involves $2 \sqrt{n_{ms}}$ error correction cycles. Its LFR is given by:
\begin{align}
    R_{1s} &= 2 \sqrt{n_{ms}} R_{ss}
\end{align}
The time taken for this is computed as:
\begin{align}
    T_{1s} &= 6 t_g \sqrt{n_{ms}}
\end{align}

For every Bell-pair Charlie Inc. creates, he keeps the surface code patch containing one half of the Bell-pair to himself, while he transports the other half to an end-node.

\textbf{Transport:} During transport, the surface code undergoes $\frac{T_{3s}}{6 t_g}$ cycles, where $T_{3s}$ is the transport time. The LFR during this time is given by:
\begin{align}
    R_{3s}=\frac{T_{3s}}{6 t_g} R_{ss}
\end{align}

\textbf{Measurement:} The surface code patches are measured as soon as the truck arrives at the end-node qATM. The time taken to measure a surface code patch is $6 t_g$, and the measurement error is $R_{ss}$. Since each surface code has just one qubit, we can measure each of these surface codes together in parallel. Since our required bit-rate is $r_e$, we can measure $r_e$ number of surface codes every second in parallel to obtain the required bandwidth. 

Now, we have the current net LFR after distribution:
\begin{align}
    R_{nets} = R_{1s} + R_{3s} + R_{ss}
\end{align}
And the total time taken for this is:
\begin{align}
    T_{tots} = T_{1s} + T_{3s} + 6 t_g
\end{align}

\textbf{Storage and Measurement:} Meanwhile, Charlie Inc., on his side, needs to store his halves of the Bell-pairs, at least until the other halves are unloaded i.e. until $T_{tots}$ amount of time has passed. The LFR for this is given by:
\begin{align}
    R_{stores} = 2 \frac{T_{tots}}{6 t_g} R_{ss}
\end{align}
The factor 2 comes from the fact that Charlie Inc. is storing two halves, each corresponding to that of Alice and Bob respectively.

Next, Charlie Inc. performs a Bell-measurement between the two surface code patches. The LFR for this is given by:
\begin{align}
    R_{BMs} = 4 \sqrt{n_{ms}} R_{ss}
\end{align}

\textbf{Overall:} The total LFR for the whole protocol is:
\begin{align}
    R_{tots} = R_{BMs} + R_{stores} + 2 R_{nets}
\end{align}

\textbf{Logistics:} Since each logical qubit can be measured in parallel, the number of patches we need per second is equal to the desired bandwidth $r_e$. Considering $S$ number of end-nodes and trucks or train cars with capacity of $n_t$ physical qubits, the number of trucks per second is given as $N_{trucks} = \frac{S r_e}{\left(\frac{n_t}{n_{ms}}\right)}$. The life cycle of a truck is the total time $T_{tots}$ from loading to unloading plus the time $T_{3s}$ needed to go back. Therefore, the total number of trucks required for the network is given by:
\begin{align}
    N_{truck_{tot_s}} = \frac{S r_e}{\left(\frac{n_t}{n_{ms}}\right)} (T_{tots} + T_{3s})
\end{align}

\begin{figure*}
\begin{minipage}{\linewidth}
\begin{algorithm}[H]
\caption{Secure Delayed Choice Quantum Network Protocol With Photonic Measurements}\label{alg:dcqn2}
\begin{algorithmic}[1]
\ForAll{end-users}
    \State Charlie Inc. prepares Bell pairs encoded in surface codes
    \State Charlie Inc. loads each half of the Bell pairs into separate sets of qLDPC code blocks via teleportation
    \State Charlie Inc. stores one set of qLDPC blocks and transports the other set to destination
    \State End-user puts in a purchase order for $n_{bits}$ number of bits as per their requirement
    \For{i from 1 to $n_{bits}$}
        \State qATM at the destination unloads a logical qubit from a qLDPC block onto a surface code  
        \State qATM converts surface code physical qubits to photons in a pre-defined order via pulsed excitation and de-excitation
        \State Photons sequentially sent to end-user's smartphone and measured
        \State Photonic measurements determine the logical surface code measurement, leading to a resultant measured bit
        \State Bit stored in the end-user's smartphone for later use.
    \EndFor
\EndFor
\State Publish measurement bases
\For{users Alice and Bob wishing to connect}
    \State Alice and Bob decide to correlate $n_{key}$ number of bits
    \For{i from 1 to $n_{key}$}
        \State Charlie Inc. unloads two corresponding qLDPC qubits onto two surface code patches
        \State Charlie Inc. performs Bell measurement on surface code patches 
        \State Charlie Inc. broadcasts the measurement result to Alice and Bob, resulting in a correlated key bit
    \EndFor
    \State Alice and Bob perform standard reconciliation techniques and get the final secure key
\EndFor
\end{algorithmic}
\end{algorithm}
\end{minipage}
\end{figure*}

\end{document}